\documentclass[aps,superscriptaddress,amsmath,amssymb,twocolumn,showpacs,floatfix,english,noshowpacs]{revtex4}

\usepackage{url}
\usepackage{bm,bbm}
\usepackage{graphicx}
\usepackage{color}
\usepackage[colorlinks=true, urlcolor=blue, linkcolor=blue, citecolor=blue, pdftex]{hyperref}
\usepackage[english]{babel}

\usepackage{soul}
\usepackage{blindtext}

\newcommand{\dagga}{{\phantom{\dagger}}}

\begin{document}

\title{Dynamical structure factor of the $J_1-J_2$ Heisenberg model on the triangular lattice: magnons, spinons, and gauge fields}

\author{Francesco Ferrari}
\affiliation{SISSA-International School for Advanced Studies, Via Bonomea 265, I-34136 Trieste, Italy}
\author{Federico Becca}
\affiliation{Dipartimento di Fisica, Universit\`a di Trieste, Strada Costiera 11, I-34151 Trieste, Italy}

\date{\today}

\begin{abstract}
Understanding the nature of the excitation spectrum in quantum spin liquids is of fundamental importance, in particular for the
experimental detection of candidate materials. However, current theoretical and numerical techniques have limited capabilities, 
especially in obtaining the dynamical structure factor, which gives a crucial characterization of the ultimate nature of the quantum
state and may be directly assessed by inelastic neutron scattering. In this work, we investigate the low-energy properties of the 
$S=1/2$ Heisenberg model on the triangular lattice, including both nearest-neighbor $J_1$ and next-nearest-neighbor $J_2$ super-exchanges,
by a dynamical variational Monte Carlo approach that allows accurate results on spin models. For $J_2=0$, our calculations are compatible 
with the existence of a well-defined magnon in the whole Brillouin zone, with gapless excitations at $K$ points (i.e., at the corners of 
the Brillouin zone). The strong renormalization of the magnon branch (also including roton-like minima around the $M$ points, i.e., 
midpoints of the border zone) is described by our Gutzwiller-projected state, where Abrikosov fermions are subject to a non-trivial 
magnetic $\pi$-flux threading half of the triangular plaquettes. When increasing the frustrating ratio $J_2/J_1$, we detect a progessive 
softening of the magnon branch at $M$, which eventually becomes gapless within the spin-liquid phase. This feature is captured by 
the band structure of the unprojected wave function (with $2$ Dirac points for each spin component). In addition, we observe an 
intense signal at low energies around the $K$ points, which cannot be understood within the unprojected picture and emerges only when 
the Gutzwiller projection is considered, suggesting the relevance of gauge fields for the low-energy physics of spin liquids.
\end{abstract}

\maketitle

\section{Introduction}

The antiferromagnetic Heisenberg model for $S=1/2$ spins interacting on the triangular lattice represents the simplest example in which 
quantum fluctuations give rise to strong modifications of the classical picture, where the minimum energy configuration shows $120^\circ$ 
order. Indeed, this was the first microscopic model that has been proposed for the realization of the so-called resonating valence-bond 
state~\cite{anderson1973,fazekas1974}. Within this approach, the ground state is described by a superposition of an exponentially large 
number of singlet coverings of the lattice, generalizing the concept of resonance introduced and developed by Rumer~\cite{rumer1932} and
Pauling~\cite{pauling1933} to describe the chemical bond. Even though recent numerical investigations~\cite{capriotti1999,chernyshev2007} 
have shown that the ground state possesses a finite magnetization in the thermodynamic limit, the results confirmed large deviations 
from classical and semiclassical limits. In addition, small perturbations on top of the nearest-neighbor Heisenberg model have shown to 
drive the system into magnetically disordered phases~\cite{zhu2018,iaconis2018}. By keeping the spin SU(2) symmetry, a natural way to 
induce further magnetic frustration is to include a next-nearest-neighbor super-exchange coupling, leading to the following 
Hamiltonian:
\begin{equation}\label{eq:hamj1j2}
{\cal H} = J_1 \sum_{\langle i,j \rangle} {\bf S}_i \cdot {\bf S}_j +
J_2 \sum_{\langle\langle i,j \rangle\rangle} {\bf S}_i \cdot {\bf S}_j,
\end{equation}
where $\langle \dots \rangle$ and $\langle \langle \dots \rangle\rangle$ indicate nearest-neighbor and next-nearest-neighbor sites
in the triangular lattice; ${\bf S}_i=(S_i^x,S_i^y,S_i^z)$ is the spin-$1/2$ operator at the site $i$ and, finally, $J_1$ and $J_2$ 
are the antiferromagnetic coupling constants. This model has been intensively investigated in the past, from the semi-classical 
approaches of the early days~\cite{jolicoeur1990,chubukov1992} to the recent numerical approaches~\cite{zhu2015,hu2015,iqbal2016}.
The latter ones indicated a rather fragile $120^\circ$ magnetic order, which is melted for $J_2/J_1 \approx 0.07(1)$ (a value that
is in very good agreement among these calculations). For larger values of the frustrating ratio $J_2/J_1$ the nature of the 
non-magnetic phase is not settled down, with evidences for either a gapped~\cite{zhu2015,hu2015} or a gapless~\cite{iqbal2016} 
spin liquid. 

An important information about the physical properties is given by the features of the low-energy spectrum. In particular, the dynamical 
structure factor $S({\bf q},\omega)$ gives a direct probe to assess the nature of the relevant excitations. These can be divided in two 
broad classes: standard gapless magnons (or gapped triplons), which exist in magnetically ordered phases (or valence-bond solids), and 
more exotic (gapped or gapless) spinons, which exist in deconfined spin liquids. In addition to spinons, another kind of excitation is 
present, due to the emergence of gauge fluctuations in the low-energy effective theory of spin liquids~\cite{savary2016}.

\begin{figure}
\includegraphics[width=\columnwidth]{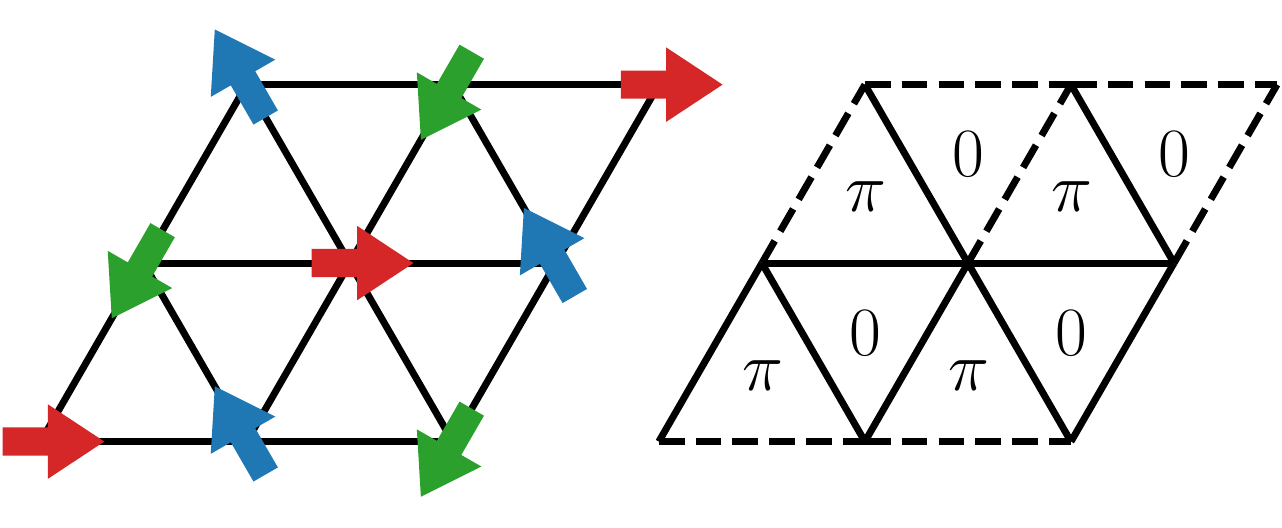}
\includegraphics[width=0.4\columnwidth]{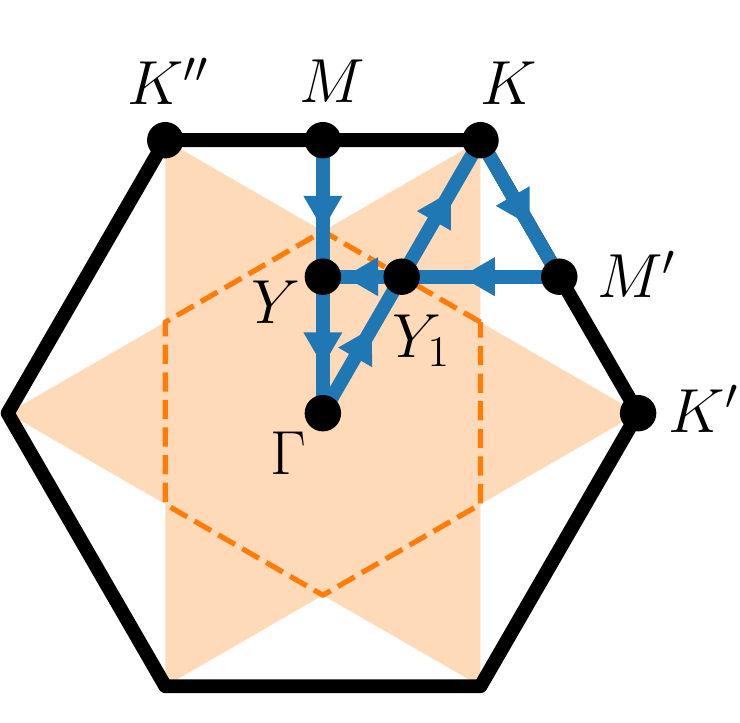}\hspace{0.05\columnwidth}
\includegraphics[width=0.4\columnwidth]{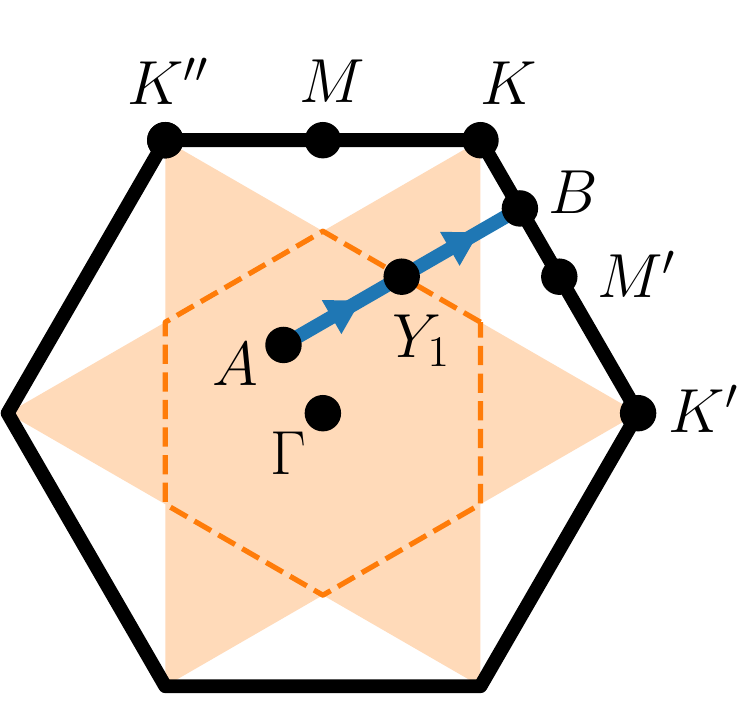}
\caption{\label{fig:latt}
Upper-left panel: the classical spin configuration (in the $XY$ plane) that is determined by the fictitious magnetic field $h$ in 
the Hamiltonian~(\ref{eq:auxham}) with ${\bf Q}=(2\pi/3,2\pi/\sqrt{3})$. Upper-right panel: pattern for the sign structure of the 
nearest-neighbor hopping $s_{i,j}$ of Eq.~(\ref{eq:auxham}), $s_{i,j}=+1$ ($-1$) for solid (dashed) lines; notice the the amplitude 
for the kinetic terms is chosen to be $t>0$. Lower-left panel: the path in the Brillouin zone that is used to plot the results of 
the dynamical structure factor of the $30\times30$ triangular lattice (blue arrows), see Figs.~\ref{fig:not}, \ref{fig:AF120}, 
\ref{fig:dispersions}, \ref{fig:007}, and~\ref{fig:125}. Lower-right panel: the path in the Brillouin zone that is used to plot the 
dynamical structure factor of the $84 \times 6$ cylinder (blue arrows), see Fig.~\ref{fig:cylin}. In both lower panels
the orange shaded area corresponds to the region of the Brillouin zone in which magnon decay is predicted by the spin-wave 
approximation~\cite{chernyshev2006,chernyshev2009} and the dashed line delimits the magnetic Brillouin zone.}
\end{figure}

For the Heisenberg model with only nearest-neighbor couplings on the triangular lattice, semi-classical approaches, based upon the
large-$S$ expansion, suggested that the excitation spectrum obtained within the leading order (i.e., within the linear spin-wave 
approximation) is subjected to significant corrections when interactions between spin waves are taken into account~\cite{starykh2006}. 
This fact is mainly due to the non-collinearity of the magnetization, which allows for three-magnon interactions. Then, despite the 
presence of long-range order, the Goldstone modes are not stable but they may decay in a large part of the Brillouin zone (see 
Fig.~\ref{fig:latt}); in particular, the existence of more than one Goldstone mode, with different velocities, immediately causes that 
magnons may be kinematically unstable, decaying into two magnons with lower energy~\cite{chernyshev2006,chernyshev2009}. A detailed 
analysis, which includes interactions among spin waves, corroborated this outcome, also showing roton-like minima at $M=(0,2\pi/\sqrt{3})$ 
and symmetry-related points (i.e., midpoints of the edges of the Brillouin zone)~\cite{chernyshev2006,chernyshev2009,zhitomirsky2013}.
The latter aspect shares similarities with the Heisenberg model on the square lattice, where minima of the magnon dispersion are present 
around $(\pi,0)$ and $(0,\pi)$~\cite{singh1995,zheng2005}. As far as the triangular lattice is concerned, aspects of the strong 
renormalization of the magnon dispersion at high energies have been confirmed by series expansions~\cite{zheng2006}. Moreover, within 
these numerical calculations, a huge downward renormalization of the one-magnon excitations is recovered, leading to a relatively 
dispersionless mode.

While there are a number of materials whose low-energy behavior can be well described by the $S=1/2$ Heisenberg model on the square 
lattice (among them, we just mention La$_2$CuO$_4$ for its relevance to cuprate superconductors~\cite{coldea2001a}), until very 
recently there were no compounds that could be well approximated by the same model on the equilateral triangular lattice. For example,
in Cs$_2$CuCl$_4$ the super-exchange couplings are not isotropic in the nearest-neighbor bonds, one out of the three being much 
stronger than the other ones (thus defining weakly-coupled zig-zag chains)~\cite{coldea2001b}. Here, inelastic neutron scattering 
measurements have shown the existence of a very broad continuum, which has been associated to spin fractionalization and spin-liquid
behavior~\cite{coldea2001b}.

Recently, measurements on Ba$_3$CoSb$_2$O$_9$ have been reported, providing evidence that it can be described by a $S=1/2$ Heisenberg 
model on the undistorted triangular lattice with predominant nearest-neighbor super-exchange couplings (a small easy-plane anisotropy 
is present, in addition to a small interlayer coupling)~\cite{shirata2012}. The initial interest was aimed at the study of the 
magnetization curve and the stabilization of magnetization plateaux~\cite{shirata2012,suzuki2013}, and the proximity to a spin liquid
phase~\cite{zhou2012}. Later, inelastic neutron scattering measurements have been performed, in order to clarify the nature of the
magnetic excitations on top of the ground state~\cite{ma2016,ito2017}. Even though Ba$_3$CoSb$_2$O$_9$ possesses long-range magnetic 
order (with $120^{\circ}$ ordering), several aspects of the magnon dispersion and the multi-magnon continuum reveal an unconventional
behavior, which can only be partly explained within semi-classical approaches. First of all, at low-energies, the magnon dispersion 
is strongly renormalized with respect to the linear spin-wave approximation; an anomalous line broadening has also been detected, 
leading to the conclusion that magnon decay may be plausible; finally, the continuum presents unexpected dispersive features at high 
energies. It should be noticed that, since neutron scattering data are sensitive to the full dynamical spin structure factor, three
copies of the magnon dispersion (translated by the ordering vectors) are visible in the spectrum. Experimental investigations have 
been also performed to infer the nature of the magnon excitations on top of the gapped phase that is stabilized at the one-third 
magnetization plateau~\cite{kamiya2018}. In this case, the situation seems to be more conventional, with the experimental results 
in relatively good agreement with theoretical predictions.

Motivated by these experimental findings, there have been a few attempts to investigate the Heisenberg model (also including small
perturbations) with both analytical and numerical tools~\cite{ghioldi2015,ghioldi2018,verresen2018,chen2018}. In particular, by using 
density-matrix renormalization group (DMRG) calculations, Verresen and collaborators~\cite{verresen2018} claimed that the magnon 
decay does not take place, because of the strong coupling interactions between quasi-particles (i.e., magnons) in the Heisenberg 
model~\cite{noteverre}. As a result of the avoided decay, the midpoint of the edge of the {\it magnetic} Brillouin zone (dubbed 
$Y_1$) displays a minimum of the magnon dispersion, possibly explaining the high-energy features seen around the $M$ point in 
Ref.~\cite{ito2017}.

Within this context, also the discovery of YbMgGaO$_4$~\cite{li2015} and, more recently, NaYbO$_2$~\cite{ding2019} will give a further 
impetus to study (generalized) spin models on the triangular lattice. In both cases, no signatures of magnetic order appear down to 
very low temperatures, suggesting the existence of a quantum spin liquid. While both materials host effective $J=1/2$ spin degrees of 
freedom, the actual low-energy Hamiltonian may be more complicated than the $SU(2)$-invariant one of Eq.~(\ref{eq:hamj1j2}); still, the 
physical properties can share many similarities with the ground state of the $J_1-J_2$ model, as suggested in Ref.~\cite{zhu2018}.

In this work, we employ a dynamical variational Monte Carlo approach~\cite{li2010} to compute the out-of-plane dynamical spin structure 
factor for the Heisenberg model on the triangular lattice, also in presence of a next-nearest-neighbor coupling $J_2$. First of all, we 
focus our attention on the model with $J_2=0$ for which we confirm huge corrections from the linear spin-wave calculations. Our results
support the idea that the magnon excitations are stable in the whole Brillouin zone; indeed, even though a {\it discrete} set of
excitations is obtained within our numerical method, the lowest-energy state for each momentum ${\bf q}$ appears to be rather well 
separated from the rest of the spectrum at higher energies, suggesting the existence of a faint continuum just above the magnon branch.
The second part of this work deals with the $J_1-J_2$ model, to highlight the modifications in the dynamical structure factor that take 
place when entering the spin-liquid phase (which, according to our variational approach, is gapless~\cite{iqbal2016}). Here, the spectrum 
shows gapless excitations at $M$ points; in addition, a strong signal at low energies is present in correspondence of the corners of the 
Brillouin zone, i.e., $K=(2\pi/3,2\pi/\sqrt{3})$ and $K^\prime=(4\pi/3,0)$. While the former aspect can be easily understood by inspecting 
the non-interacting spinon band structure, the latter one is a genuine feature that emerges from the Gutzwiller projector, which includes 
interactions between spinons and gauge fields. Indeed, while the non-interacting wave function corresponds to a mean-field approximation, 
in which gauge fields are completely frozen, the Gutzwiller projection has the effect of inserting back the temporal fluctuations of 
those fields~\cite{wenbook}. In this respect, it is worth mentioning that a recent field-theoretical analysis indicated the existence of 
low-energy (triplet) monopole excitations at the zone corners, which are expected to contribute to the dynamical structure 
factor~\cite{song2018}.

\section{Dynamical variational Monte Carlo}\label{sec:method}

The dynamical structure factor, which is directly measured within inelastic neutron scattering experiments, can be used to unveil the 
nature of the elementary excitations of the models/materials under investigation. In its spectral form, this quantity reads as
\begin{equation}\label{eq:dsf}
S^{a}({\bf q},\omega) = \sum_{\alpha} |\langle \Upsilon_{\alpha}^q | S^{a}_q | \Upsilon_0 \rangle|^2 \delta(\omega-E_{\alpha}^q+E_0),
\end{equation}
where $|\Upsilon_0\rangle$ and $\{|\Upsilon_{\alpha}^q\rangle\}_\alpha$ are the ground state and the set of all excited states with 
momentum $q$, whose corresponding energies are $E_0$ and $\{E_{\alpha}^q\}_{\alpha}$, respectively. In this work, we evaluate the dynamical 
structure factor of the spin model~(\ref{eq:hamj1j2}) by directly constructing accurate variational {\it Ansatze} for its ground state and 
a few low-energy excited states. Our variational approach is based on the so-called {\it parton} construction, in which the spin degrees of 
freedom of the model are rewritten in terms of auxiliary fermionic operators~\cite{savary2016,wen2002}. The fermionic language constitute a 
versatile framework to define variational wave functions for both magnetically ordered and disordered phases of matter. The present Section 
is dedicated to the introduction of the fermionic wave functions for spin models and to the description of the variational Monte Carlo 
method employed for the calculation of the dynamical structure factor.

\subsection{Gutzwiller-projected fermionic wave functions for the ground state}\label{sec:wavefunctions}

Here, for the sake of generality, we consider a generic $SU(2)$ model for frustrated spin systems, which consists of a set of spin-$1/2$ 
degrees of freedom sitting on the sites of a lattice and interacting through the Heisenberg exchange couplings $J_{i,j}$:
\begin{equation}\label{eq:generic_heis}
{\cal H} = \sum_{i,j} J_{i,j} {\bf S}_i \cdot {\bf S}_j.
\end{equation}
The interplay of the different interactions can lead to the stabilization of different phases of matter. In absence of frustration, i.e.,
when no competing couplings are present, the ground state may develop some kind of magnetic order, which minimizes the classical energy of 
the model. On the contrary, when different interactions compete with each other, magnetically disordered phases can arise, such as spin 
liquids. 

The first attempt to describe spin-liquid states dates back to the resonating valence-bond approach, where a variational wave function is
defined in terms of a linear superposition of singlet coverings of the lattice~\cite{anderson1973}. More recently, Wen~\cite{wen2002} 
developed a general approach to classify and construct spin-liquid states, which satisfy all the symmetries of a given lattice model. 
This method is built upon the introduction of auxiliary Abrikosov fermions, which form a projective representation of $S=1/2$ spin operators: 
\begin{equation}\label{eq:Sabrikosov}
{\bf S}_i = \frac{1}{2} \sum_{\alpha,\beta} c_{i,\alpha}^\dagger 
\boldsymbol{\sigma}_{\alpha,\beta} c_{i,\beta}^\dagga.
\end{equation}
Here $c_{i,\alpha}^\dagga$ ($c_{i,\alpha}^\dagger$) destroys (creates) a fermion with spin $\alpha=\uparrow,\downarrow$ on site $i$, and the 
vector $\boldsymbol{\sigma}=(\sigma_x,\sigma_y,\sigma_z)$ is the set of Pauli matrices. The anticommutation relations among fermions ensure 
that the Abrikosov representation yields the correct commutation relations among different spin components. Still, in order to faithfully 
reproduce the Hilbert space of the original spin model, only configurations with one fermion per site must be considered, which implies that 
the Abrikosov fermions must satisfy the constraint:
\begin{equation}\label{eq:Gutz_constraint1}
c^\dagger_{i,\uparrow}c^\dagga_{i,\uparrow}+c^\dagger_{i,\downarrow}c^\dagga_{i,\downarrow}=1,
\end{equation}
or equivalently:
\begin{equation}\label{eq:Gutz_constraint2}
c^\dagger_{i,\uparrow} c^\dagger_{i,\downarrow}=0,
\end{equation}

Besides constant terms, the Hamiltonian of Eq.~(\ref{eq:generic_heis}) can be rewritten in terms of Abrikosov fermions as follows:
\begin{equation}\label{eq:quartic_ham}
 {\cal H} = -\frac{1}{2}\sum_{i,j} \sum_{\alpha,\beta} J_{i,j} \left(
 c_{i,\alpha}^\dagger c_{j,\alpha}^\dagga c_{j,\beta}^\dagger c_{i,\beta}^\dagga 
 + \frac{1}{2} c_{i,\alpha}^\dagger c_{i,\alpha}^\dagga c_{j,\beta}^\dagger c_{j,\beta}^\dagga \right).
\end{equation}
At this stage, the Hamiltonian~(\ref{eq:quartic_ham}) with the constraints of Eqs.~(\ref{eq:Gutz_constraint1}) and~(\ref{eq:Gutz_constraint2}) 
give an {\it exact} representation of the original model. In order to tackle the above interacting fermionic system, one possibility is to 
perform a mean-field decoupling~\cite{wen2002}. For the purpose of studying spin-liquid phases, we keep only the mean-field terms that do not 
break the $SU(2)$ symmetry of the original spins. The result is a quadratic Hamiltonian:
\begin{eqnarray}\label{eq:generic_mf}
 {\cal H}_{0} = \sum_{i,j} \sum_{\sigma} t_{i,j} c_{i,\sigma}^\dagger c_{j,\sigma}^\dagga +
 \sum_{i,j}  \Delta_{i,j} c_{i,\uparrow}^\dagger c_{j,\downarrow}^\dagger + h.c. \nonumber \\
 +\sum_{i} \sum_{\sigma} \mu_i c_{i,\sigma}^\dagger c_{i,\sigma}^\dagga +
 \sum_{i} \zeta_{i} c_{i,\uparrow}^\dagger c_{i,\downarrow}^\dagger + h.c.,
\end{eqnarray}
which contains a hopping term $t_{i,j}$ and a singlet pairing term $\Delta_{i,j}$, which are related to the expectation values 
$\langle c_{j,\sigma}^\dagger c_{i,\sigma}^\dagga\rangle$ and $\langle c_{i,\sigma} c_{j,-\sigma}\rangle$, respectively. In addition, the 
one-fermion-per-site constraint of the parton construction is enforced in a {\it global} fashion by including a chemical potential $\mu_i$ 
and an onsite-pairing $\zeta_{i}$ as Lagrange multipliers in ${\cal H}_{0}$~\cite{wen2002}. Within the mere mean-field approach, the 
parameters of ${\cal H}_{0}$ are computed self-consistently and define a low-energy effective theory for the spin model under investigation. 
However, the ground state of ${\cal H}_{0}$, named $|\Phi_0 \rangle$, satisfies the constraints of Eqs.~(\ref{eq:Gutz_constraint1}) 
and~(\ref{eq:Gutz_constraint2}) only on average and, therefore, does not represent a valid wave function for spins. Within this approach, 
a full treatment of the original spin model requires the inclusion of all fluctuations of the parameters around the mean-field solution. 
Since this task is in general unfeasible, an alternative approach can be pursued, in which the Hamiltonian ${\cal H}_{0}$ is exploited 
as a starting point for the definition of a variational wave function for the initial spin model. Indeed, the one-fermion-per-site 
constraint can be enforced exactly by applying the Gutzwiller projector,
\begin{equation}
 \mathcal{P}_G= \prod_i (n_{i,\uparrow}-n_{i,\downarrow})^2,
\end{equation}
to the ground state wave function of ${\cal H}_{0}$. We emphasize that in general the Gutzwiller projection cannot be treated analytically, 
due to its intrinsic many-body character, however it can be considered within Monte Carlo sampling. At variance with the mean-field treatment, 
in the variational approach the parameters of ${\cal H}_{0}$ are not computed self-consistently, but are optimized in order to minimize the 
energy of the Gutzwiller-projected {\it Ansatz} $\mathcal{P}_G|\Phi_0 \rangle$.  

The artificial enlargement of the Hilbert space introduced by the parton construction gives rise to a {\it gauge redundancy} in the 
representation of the spin degrees of freedom. Specifically, the mapping~(\ref{eq:Sabrikosov}) is invariant under {\it local} $SU(2)$ 
transformations of the Abrikosov fermions operators~\cite{wen2002}. As a consequence, all physical properties of the spins are independent on 
the gauge choice for fermions. For example, whenever we perform $SU(2)$ transformations to the unprojected Hamiltonian ${\cal H}_{0}$, the
variational wave function with the Gutzwiller projector remains invariant. Exploiting this gauge redundancy, it is possible to classify all 
the quadratic Hamiltonians ${\cal H}_{0}$ whose Gutzwiller-projected ground states fulfill the symmetries of the lattice model. 
This procedure, known as projective symmetry group analysis~\cite{wen2002}, provides a recipe to construct all the distinct spin liquid 
{\it Ansatze} for a given spin model. From a variational point of view, the spin-liquid wave function with the lowest variational energy is 
the one which better describes the true ground state of the model.

In general, the variational {\it Ansatze} defined by Gutzwiller-projecting the ground state of Eq.~(\ref{eq:generic_mf}) do not display any 
magnetic order~\cite{li2013}. For the purpose of defining suitable wave functions for magnetically ordered phases, an additional term can be 
added to ${\cal H}_{0}$:
\begin{equation}\label{eq:magnfield}
 {\cal H}_{0} \mapsto {\cal H}_{0} + h \sum_{i}  
             \left ( e^{i \mathbf{Q} \cdot \mathbf{R}_i} c_{i,\uparrow}^\dagger c_{i,\downarrow}^\dagga
            + e^{-i \mathbf{Q} \cdot \mathbf{R}_i} c_{i,\downarrow}^\dagger c_{i,\uparrow}^\dagga \right ).
\end{equation}
Here, $h$ is a {\it fictitious} magnetic field which lies in the $XY$ plane and displays a periodic pattern defined by the pitch vector 
$\mathbf{Q}$. Since the ground-state wave function of the Hamiltonian~(\ref{eq:magnfield}) tends to overestimate the magnetic 
order~\cite{becca2011}, further transverse quantum fluctuations are added through the application of a spin-spin Jastrow factor,  
\begin{equation}
\mathcal{J}_s=\exp \left ( \frac{1}{2} \sum_{i,j} v_{i,j} S^z_i S^z_j \right ),
\end{equation}
to the Gutzwiller-projected state. Specifically, the complete form of the variational wave functions employed in this work is
\begin{equation}\label{eq:wf}
|\Psi_0\rangle= \mathcal{P}_{S_z} \mathcal{J}_s \mathcal{P}_G |\Phi_0 \rangle,
\end{equation}
where in addition to the Gutzwiller projection and the Jastrow factor, we apply a projector enforcing zero value for the $z$-component of the 
total spin ($\mathcal{P}_{S_z}$).

By using this approach, the variational phase diagram for the $J_1-J_2$ model on the triangular lattice has been obtained in Ref.~\cite{iqbal2016}:
the system undergoes a phase transition between a magnetically ordered phase to a gapless spin liquid at ${J_2/J_1 \approx 0.08}$. For this 
model, the optimal variational wave functions are obtained by considering only a hopping term (no pairing) 
and the fictitious magnetic field in the quadratic Hamiltonian:
\begin{eqnarray}\label{eq:auxham}
\mathcal{H}_0 &=& t \sum_{\langle i,j \rangle} s_{i,j} c_{i,\sigma}^\dagger c_{j,\sigma}^\dagga \nonumber \\
&+& h \sum_{i}  \left ( e^{i \mathbf{Q} \cdot \mathbf{R}_i} c_{i,\uparrow}^\dagger c_{i,\downarrow}^\dagga
+ e^{-i \mathbf{Q} \cdot \mathbf{R}_i} c_{i,\downarrow}^\dagger c_{i,\uparrow}^\dagga \right ).
\end{eqnarray}
Here $t$ is a first-neighbor hopping with a non-trivial sign structure ($s_{i,j} = \pm 1$) which generates a pattern of alternating $0$ 
and $\pi$ fluxes through the triangular plaquettes of the lattice, see Fig.~\ref{fig:latt}; $h$ is a fictitious magnetic field which displays 
the classical $120^\circ$ order with ${\bf Q}=(2\pi/3,2\pi/\sqrt{3})$, see Fig.~\ref{fig:latt} (considering ${\bf Q}=(4\pi/3,0)$ would not 
change the physical content of the ground state wave function). All the parameters included in $\mathcal{H}_0$ and the pseudopotential 
$v_{i,j}$ (one parameter for each distance $|{\bf R}_i-{\bf R}_j|$ in the translational invariant lattice) entering the Jastrow factor can 
be optimized to minimize the variational energy. While in the magnetic phase of the system the optimal value for the ratio $h/t$ is finite, 
for $J_2/J_1 \gtrsim 0.08$ the system enters the spin liquid phase and the magnetic field parameter vanishes in the thermodynamic 
limit~\cite{iqbal2016}. The values of the fictitious magnetic field as a function of $J_2/J_1$ can be found in Ref.~\cite{iqbal2016}.

In this work we compute the dynamical structure factor for the $J_1-J_2$ model on the $30 \times30$ triangular lattice. For $J_2=0$, we first
consider the crudest approximation for the ground state, which consists in setting the hopping term $t$ to zero. The resulting wave function 
is equivalent to the state of Ref.~\cite{huse1988} with only a two-body Jastrow factor. Much more accurate results are then obtained by restoring
the hopping term in the Hamiltonian and optimizing all the variational parameters, for the cases $J_2=0$ and $J_2/J_1=0.07$. On the other hand, 
when the system is in the spin liquid regime ($J_2/J_1=0.09$ and $J_2/J_1=0.125$), the fictitious magnetic field is vanishing and the Jastrow 
factor is not considered, because of its negligible effects on the variational results. According to the projective symmetry group classification, 
the wave function obtained by considering only the hopping term in $\mathcal{H}_0$ is a fully symmetric $U(1)$ spin liquid~\cite{lu2016}.

\subsection{Dynamical structure factor}\label{sec:dynamical}

\begin{figure}
\includegraphics[width=\columnwidth]{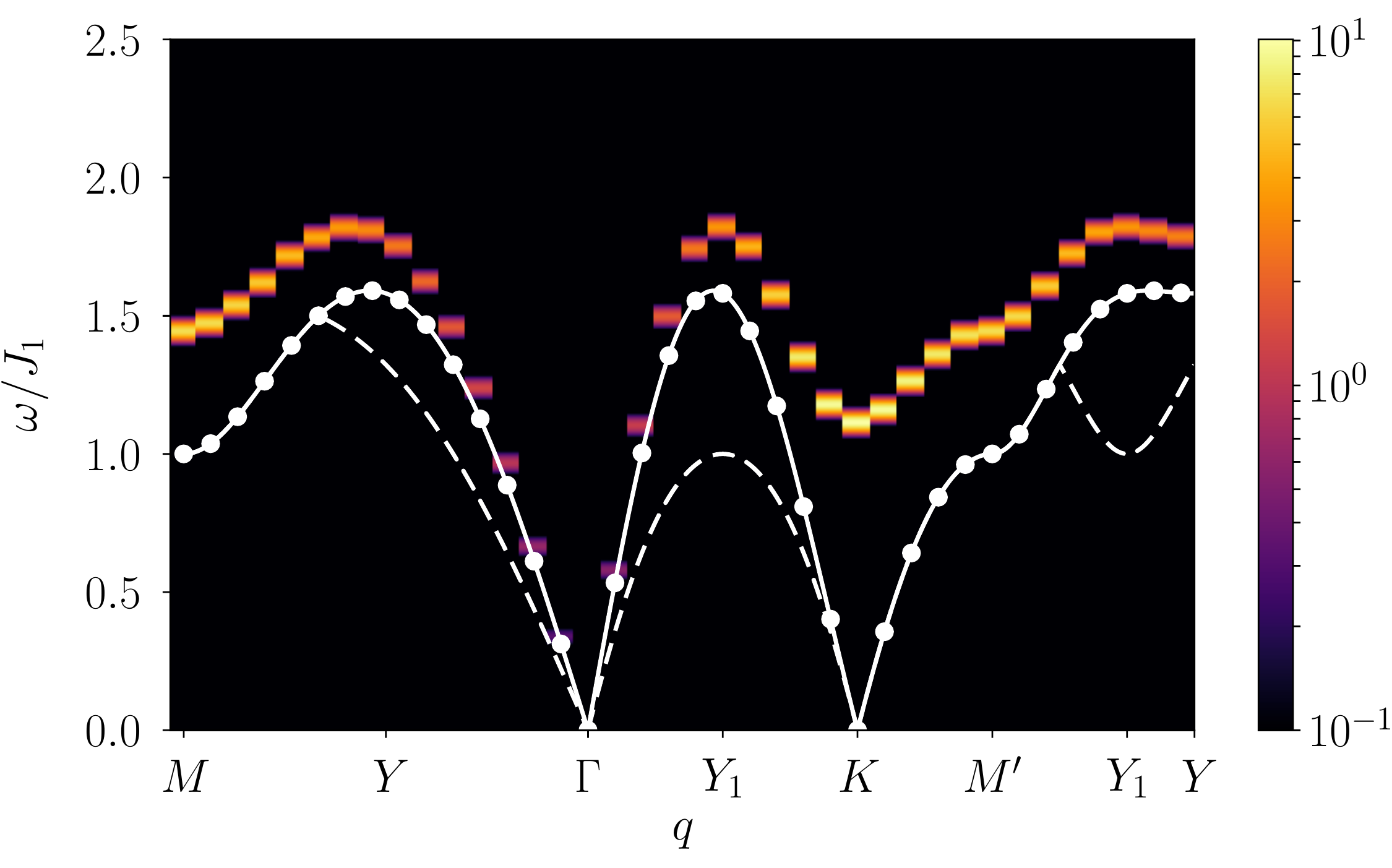}
\caption{\label{fig:not}
Dynamical structure factor of the nearest-neighbor Heisenberg model on the triangular lattice obtained by using the variational wave function 
of Eq.~(\ref{eq:wf}) and~(\ref{eq:auxham}) with $t=0$ on the $30 \times 30$ cluster. The path along the Brillouin zone is shown in 
Fig.~\ref{fig:latt}. A Gaussian broadening of the spectrum has been applied ($\sigma=0.02J_1$). The spin-wave energies of the magnon branch 
($\epsilon_q$), on the same cluster size, are represented by the white dots connected with a solid line. The dashed line corresponds to the 
bottom of the continuum within linear spin waves, i.e.  $E_q=\min_{k} \{ \epsilon_{q-k} + \epsilon_{k} \}$. Notice that $E_{q}<\epsilon_{q}$ 
in most of the Brillouin zone, as obtained in Ref.~\cite{chernyshev2006,chernyshev2009}.}
\end{figure}

\begin{figure}
\includegraphics[width=\columnwidth]{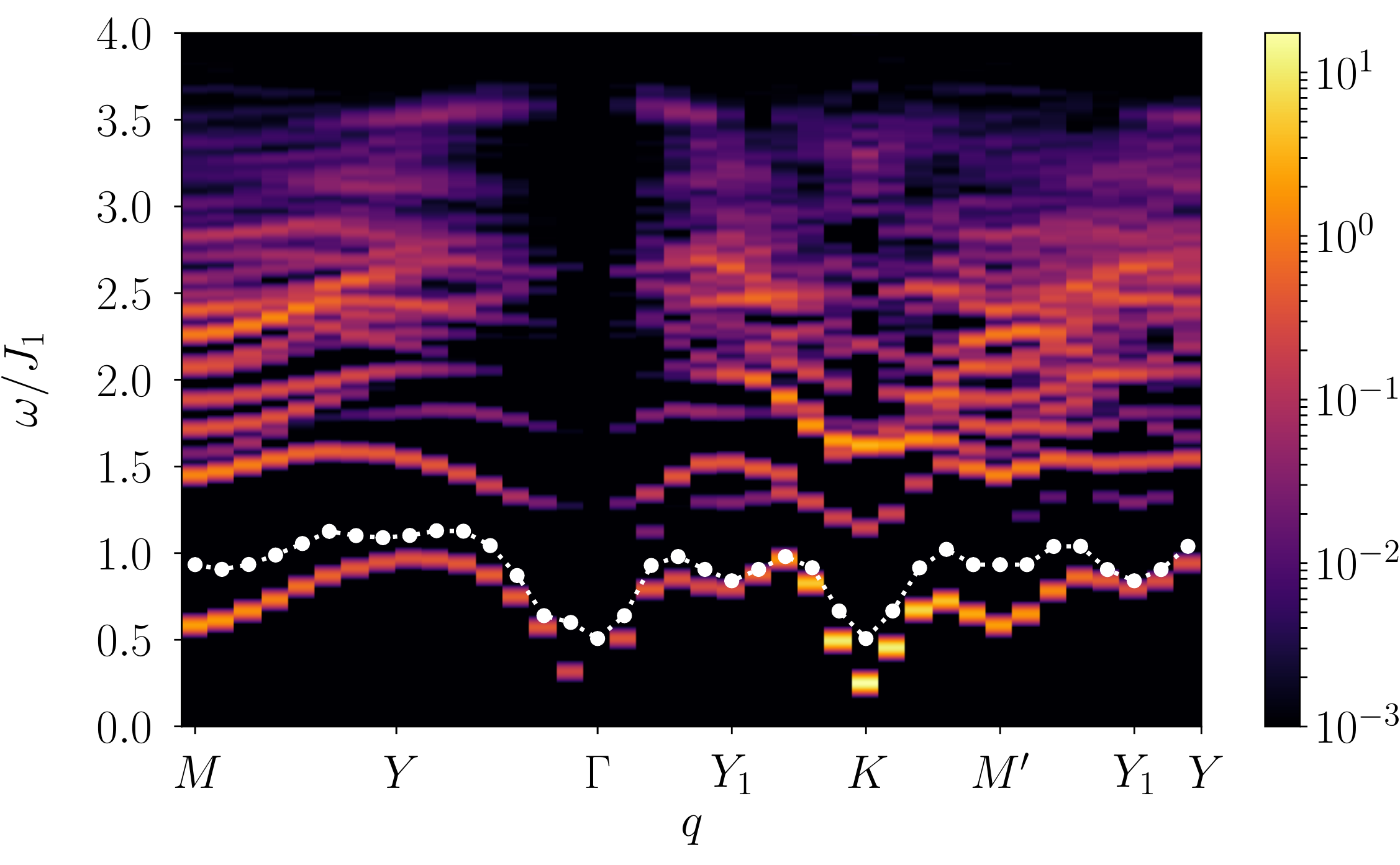}
\caption{\label{fig:AF120}
The same as Fig.~\ref{fig:not} but for the optimal variational wave function with both hopping $t$ and fictitious magnetic field $h$. 
The path along the Brillouin zone is shown in Fig.~\ref{fig:latt}. The dotted line denotes the bottom of the continuum 
$E_{q}=\min_{k} \{E_{0}^{q-k}+E_{0}^{k}\}$, where $E_{0}^{q}$ is the lowest energy for a given momentum ${\bf q}$ obtained within our 
variational approach. Since the spectrum is gapless at the $\Gamma$ point, we exclude the cases ${\bf k}=(0,0)$ and ${\bf k}={\bf q}$ 
in the search of the minimum, because the resulting $E_{q}$ would simply coincide with the energy of the magnon branch $E_{0}^{q}$ 
all over the Brillouin zone. The purpose of this kinematic analysis is to show that no magnon decay can yield an energy $E_{q}$
which is lower than the one of the magnon branch $E_{0}^{q}$ (in constrast with spin wave results).}
\end{figure}

\begin{figure}
\includegraphics[width=\columnwidth]{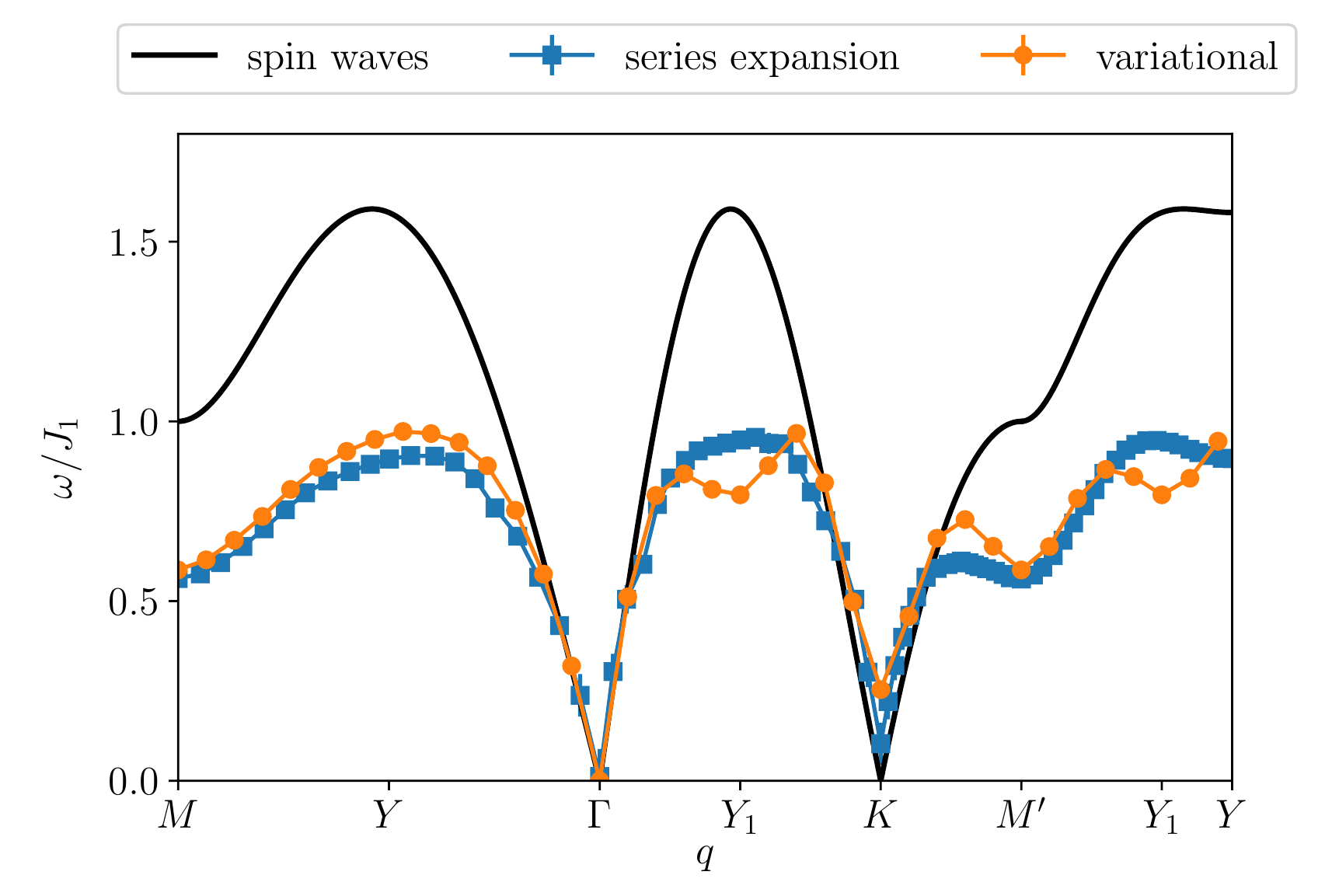}
\caption{\label{fig:dispersions}
Energies of the magnon branch for the nearest-neighbor Heisenberg model on the triangular lattice obtained with different methods. The path 
along the Brillouin zone is shown in Fig.~\ref{fig:latt}. The black line corresponds to linear spin wave, the blue squares to series 
expansion~\cite{zheng2006}, and the orange circles to our variational results (on the $30 \times 30$ cluster).}
\end{figure}

\begin{figure}
\includegraphics[width=\columnwidth]{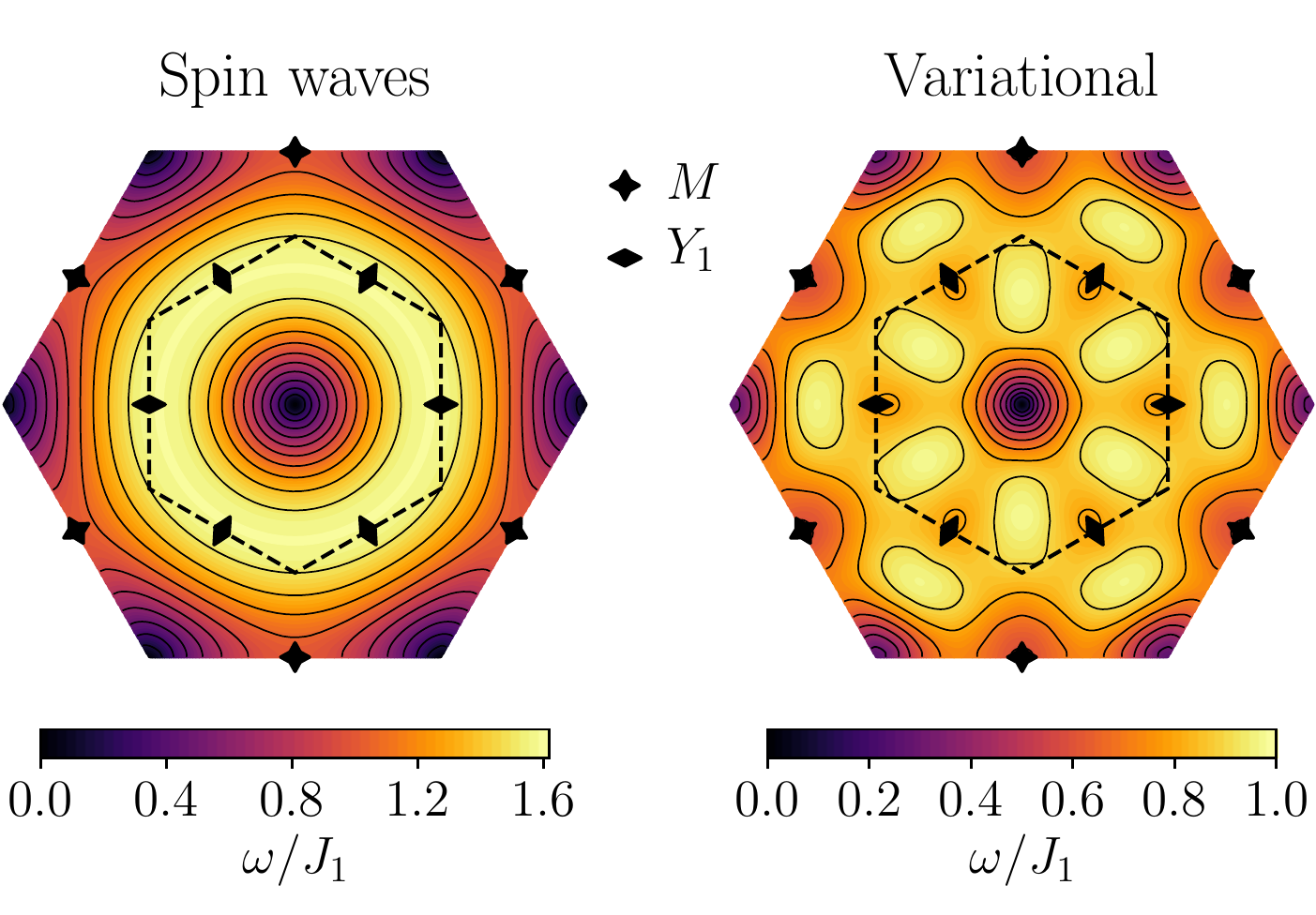}
\caption{\label{fig:dispersion}
Dispersion relation of the magnon branch (i.e., the lowest-energy excitation) as obtained within our variational approach (on the $30\times 30$ 
cluster). The linear spin-wave results are also reported for comparison. Dashed lines represent the edges of the magnetic Brillouin zone. The 
presence of the roton minima at the $M$ and $Y_1$ points in the variational spectrum is evident.}
\end{figure}

As already mentioned, the dynamical structure factor of the $J_1-J_2$ model is computed by constructing variational {\it Ansatze} to approximate 
the low-energy excited states of the system. Here we limit ourselves to the calculation of the out-of-plane component $S^z({\bf q},\omega)$, 
and we employ the technique outlined in Ref.~\cite{li2010,ferrari2018a,ferrari2018b}, which is briefly summarized in the following.

First, we find the optimal variational \textit{Ansatz} for the ground state of the model, which has the form of Eq.~(\ref{eq:wf}), by 
minimizing the variational energy. The resulting wave function is employed as a reference state to construct a set of projected particle-hole 
excitations with a given momentum $q$:
\begin{equation}\label{eq:qRstate}
|q,R\rangle =  \mathcal{P}_{S_z} \mathcal{J}_s \mathcal{P}_G 
\frac{1}{\sqrt{N}} \sum_{i}\sum_{\sigma} e^{i {\bf q} \cdot {\bf R}_i} \sigma c^\dagger_{i+R,\sigma}c^\dagga_{i,\sigma} |\Phi_0\rangle.
\end{equation}
These states are labelled by $R$, which runs over all lattice vectors. We approximate the low-energy excited states of the model by using 
linear combinations of the elements of the basis set $\{|q,R\rangle\}_R$:
\begin{equation}\label{eq:psinq}
 |\Psi_n^q\rangle=\sum_R A^{n,q}_R |q,R\rangle.
\end{equation}
For a certain momentum {\bf q}, we consider the Schr{\"o}dinger equation for the $J_1-J_2$ Hamiltonian restricting the form of its eigenvectors 
to the one of Eq.~(\ref{eq:psinq}),  i.e. ${ {\cal H}|\Psi_n^q\rangle = E_n^q |\Psi_n^q\rangle }$. Expanding everything in terms of 
$\{|q,R\rangle\}_R$, we arrive to the following generalized eigenvalue problem
\begin{equation}\label{eq:general_eig_prob}
\sum_{R^\prime} \langle q,R|{\cal H}|q,R^\prime \rangle  A^{n,q}_{R^\prime} = E_n^q \sum_{R^\prime} \langle q,R|q,R^\prime \rangle 
A^{n,q}_{R^\prime},
\end{equation}
which is solved to find the expansion coefficients $A^{n,q}_R$ and the energies $E_n^q$ of the excitations. All the matrix elements,
$\langle q,R|{\cal H}|q,R^\prime \rangle$ and $\langle q,R|q,R^\prime \rangle$, are evaluated within the Monte Carlo procedure, by 
sampling according to the variational ground-state wave function. Finally the dynamical structure factor is computed by:
\begin{equation}\label{eq:Szz_practical}
S^{z}({\bf q},\omega) = \sum_n |\langle \Psi_{n}^q | S^{z}_q | \Psi_0 \rangle|^2 \delta(\omega-E_{n}^q+E_0^{\rm var}),
\end{equation}
where $E_0^{\rm var}$ is the variational energy of $|\Psi_0 \rangle$.

\section{Results}\label{sec:results}

In this section, we present the numerical calculations for the dynamical structure factor $S({\bf q},\omega)$ obtained by the 
variational approach described in the previous section. First, we discuss the case of the Heisenberg model with only nearest-neighbor
super-exchange $J_1$, also comparing our results with recent DMRG calculations~\cite{verresen2018}. Then, we include the 
next-nearest-neighbor coupling $J_2$ to increase frustration and melt the magnetic order. In this way, a gapless spin-liquid regime 
is reached for $J_2/J_1 \approx 0.08$~\cite{iqbal2016}.

\subsection{The nearest-neighbor model with $J_2=0$}

Let us start our analysis by considering the case in which the ground-state wave function only contains the fictitious magnetic field,
i.e., $t=0$. In this case, the Abrikosov fermions are completely localized (e.g., the eigenvalues of the auxiliary Hamiltonian define 
flat bands) and the wave function corresponds to the Jastrow state of Ref.~\cite{huse1988} with only a two-body Jastrow factor. The 
results for the dynamical structure factor on the $30 \times 30$ cluster are shown in Fig.~\ref{fig:not}. Here, the spectrum consists 
of a {\it single} mode, which is identified as the magnon excitation (no continuum is visible). Notice that only one magnon branch is 
visible, related to the magnon dispersion $\epsilon_{q}$, since we consider the out-of-plane dynamical structure factor (the {\it folded}
branches $\epsilon_{q \pm K}$ do not contribute to the signal). Remarkably, the dispersion of the magnon branch is possible thanks to 
the Jastrow factor, since the wave function without it would give rise to a trivially flat (gapped) excitation spectrum, reflecting 
the non-interacting band structure of fermions. By contrast, the long-range Jastrow term is able to produce a reasonable magnon mode, 
which agrees fairly well with the spin-wave calculations. In paticular, the spectrum is gapless at $\Gamma=(0,0)$ (with a vanishingly 
small weight). Instead, in constrast to spin waves, which correctly predict gapless magnons at $K$ and $K^\prime$ due to the coplanar
$120^\circ$ order, this simple wave function leads to a gapped spectrum at the corners of the Brillouin zone. In connection to that, 
the out-of-plane static structure factor $S^z({\bf q})=\int d\omega S^z({\bf q},\omega)$ does not diverge at $K$ or $K^\prime$ when 
$L \to \infty$, showing only a maximum.

\begin{figure}
\includegraphics[width=\columnwidth]{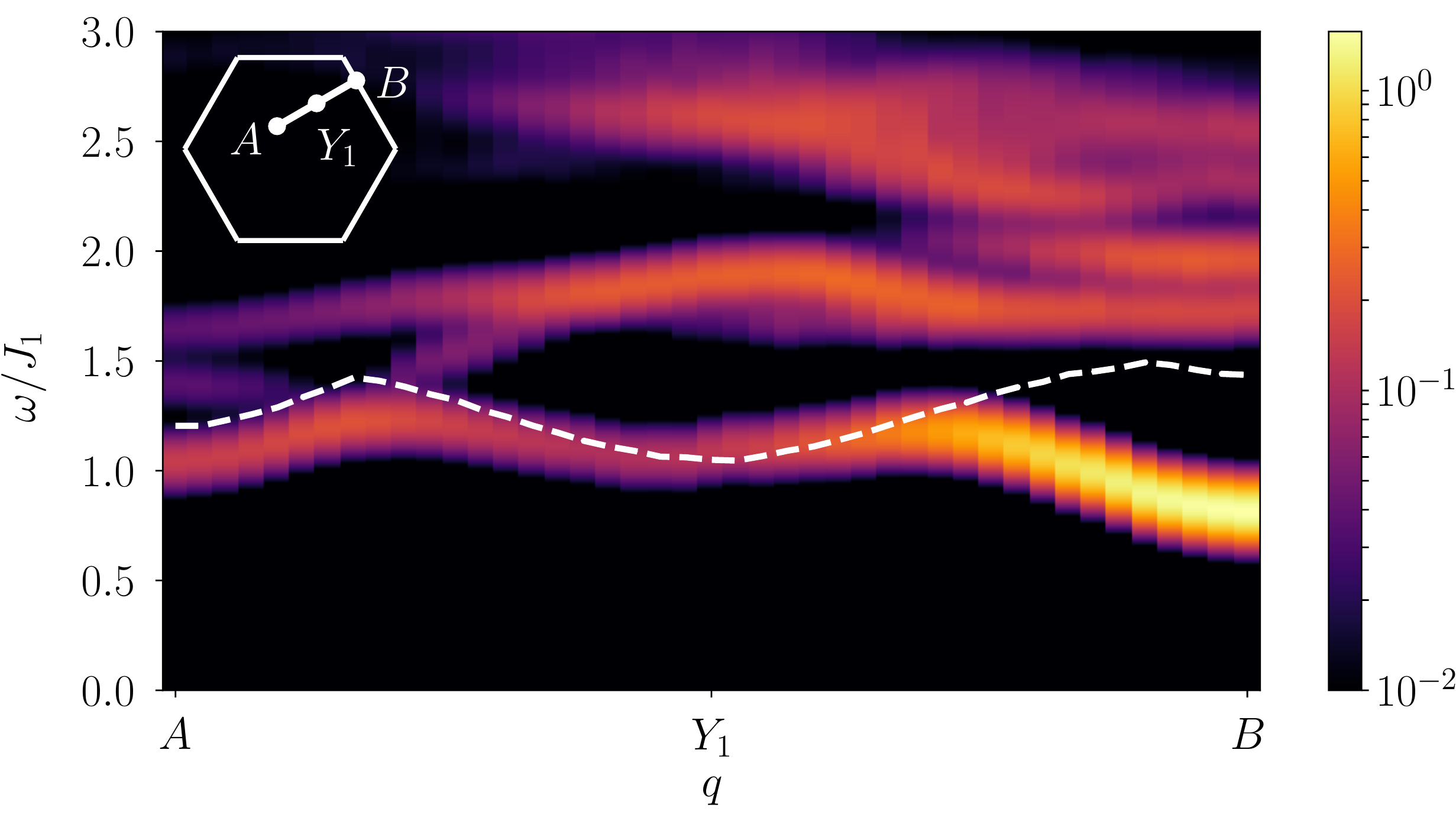}
\caption{\label{fig:cylin}
The dynamical structure factor for the nearest-neighbor Heisenberg model on a cylindrical geometry ($84 \times 6$), to make a close
comparison with DMRG calculations by Verresen and collaborators~\cite{verresen2018}. We apply a Gaussian broadening to the spectrum 
which is equivalent to the one of the aforementioned DMRG result ($\sigma=0.077J_1$). The path in the Brillouin zone is shown in the 
inset and in Fig.\ref{fig:latt} (the point $A$ lies at $1/4$ of the $\Gamma-K^{\prime\prime}$ line, where 
$K^{\prime\prime}=(-2\pi/3,2\pi/\sqrt{3})$; the point $B$ lies at $1/4$ of the $K-K^\prime$ line). The dashed line denotes the bottom 
of the continuum, which is evaluated by taking $E_{q}=\min\{E_{0}^{q-K}+E_{0}^{K},E_{0}^{q+K}+E_{0}^{-K}\}$, where $E_{0}^{q}$ is 
the lowest energy for a given momentum $q$ obtained within our variational approach and $K=(2\pi/3,2\pi/\sqrt{3})$.}
\end{figure}

\begin{figure}
\includegraphics[width=\columnwidth]{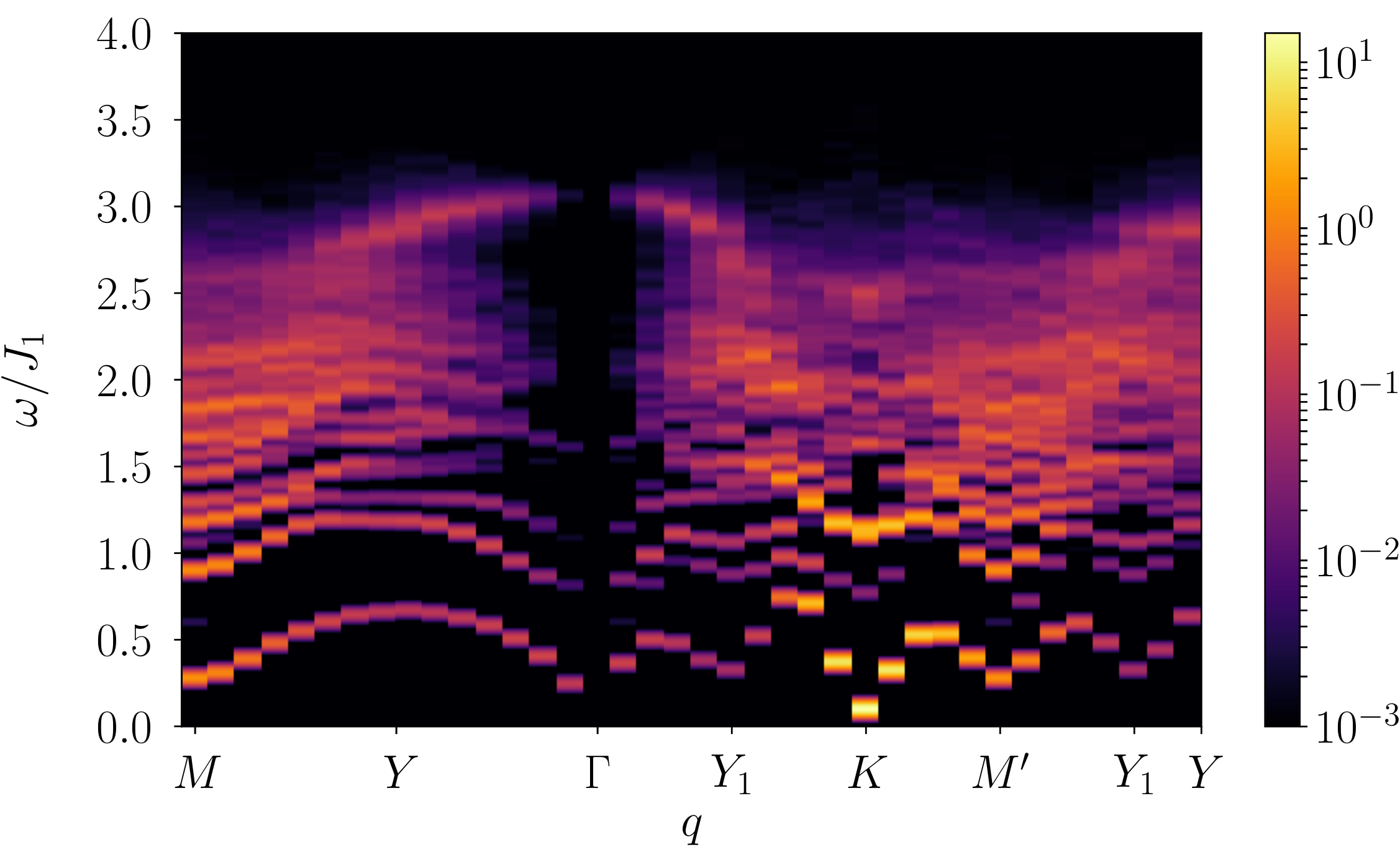}
\includegraphics[width=\columnwidth]{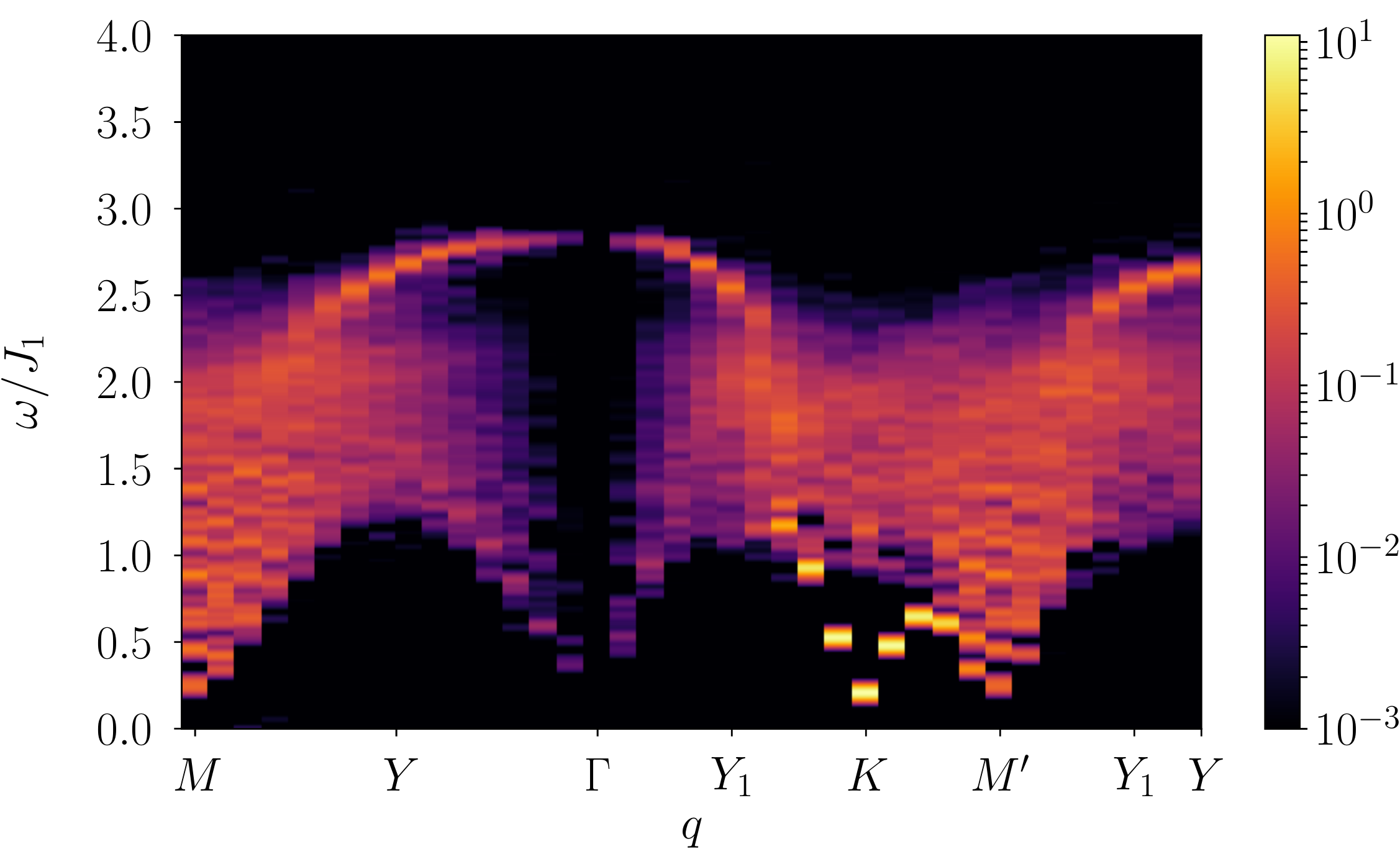}
\caption{\label{fig:007}
The dynamical structure factor for the $J_1-J_2$ Heisenberg model on the $30 \times 30$ cluster with $J_2/J_1=0.07$ (above) and
$J_2/J_1=0.09$ (below). The path along the Brillouin zone is shown in Fig.~\ref{fig:latt} and a Gaussian broadening of the spectrum 
has been applied ($\sigma=0.02J_1$).}
\end{figure}

\begin{figure*}
\includegraphics[width=\columnwidth]{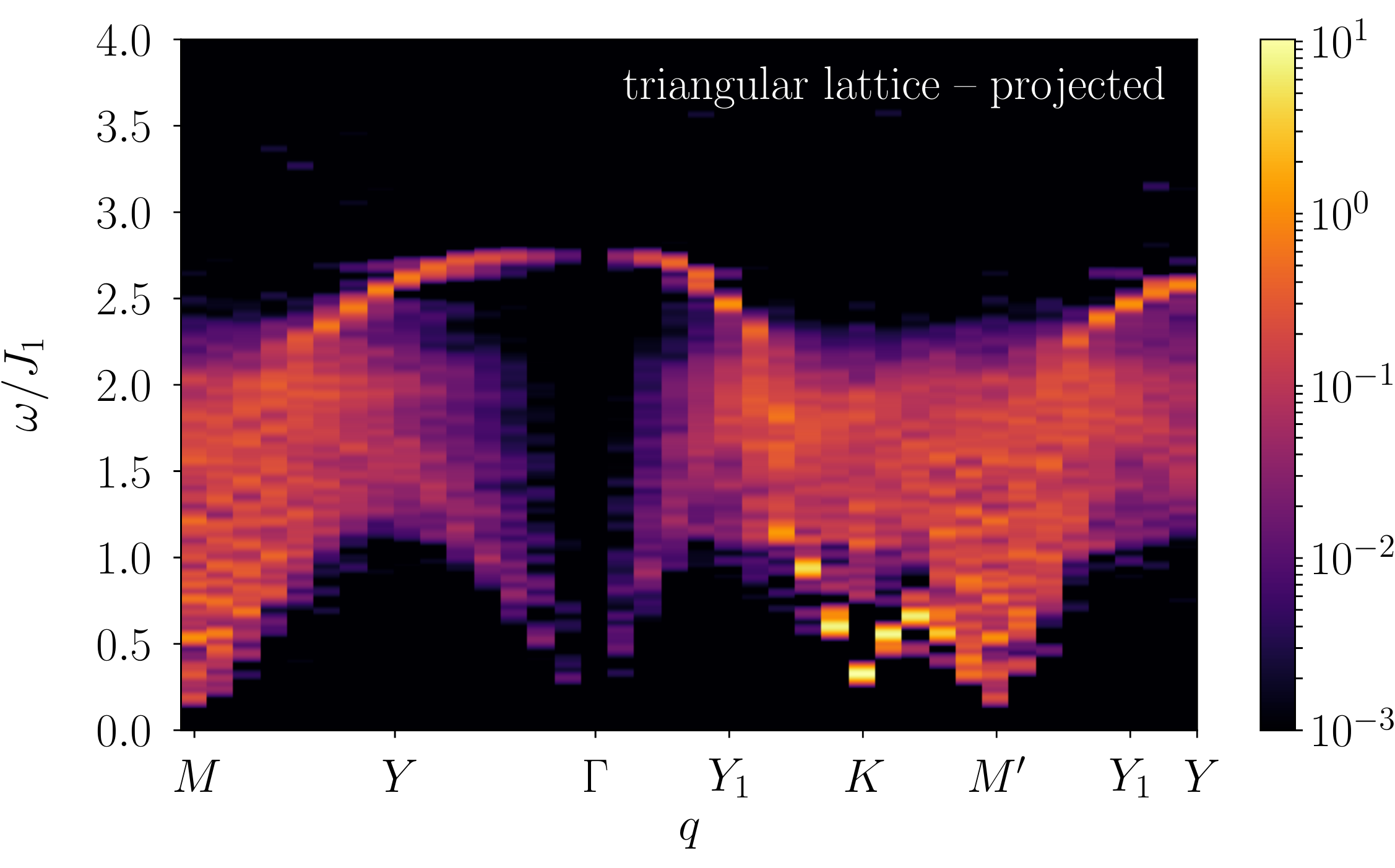}\hfill
\includegraphics[width=\columnwidth]{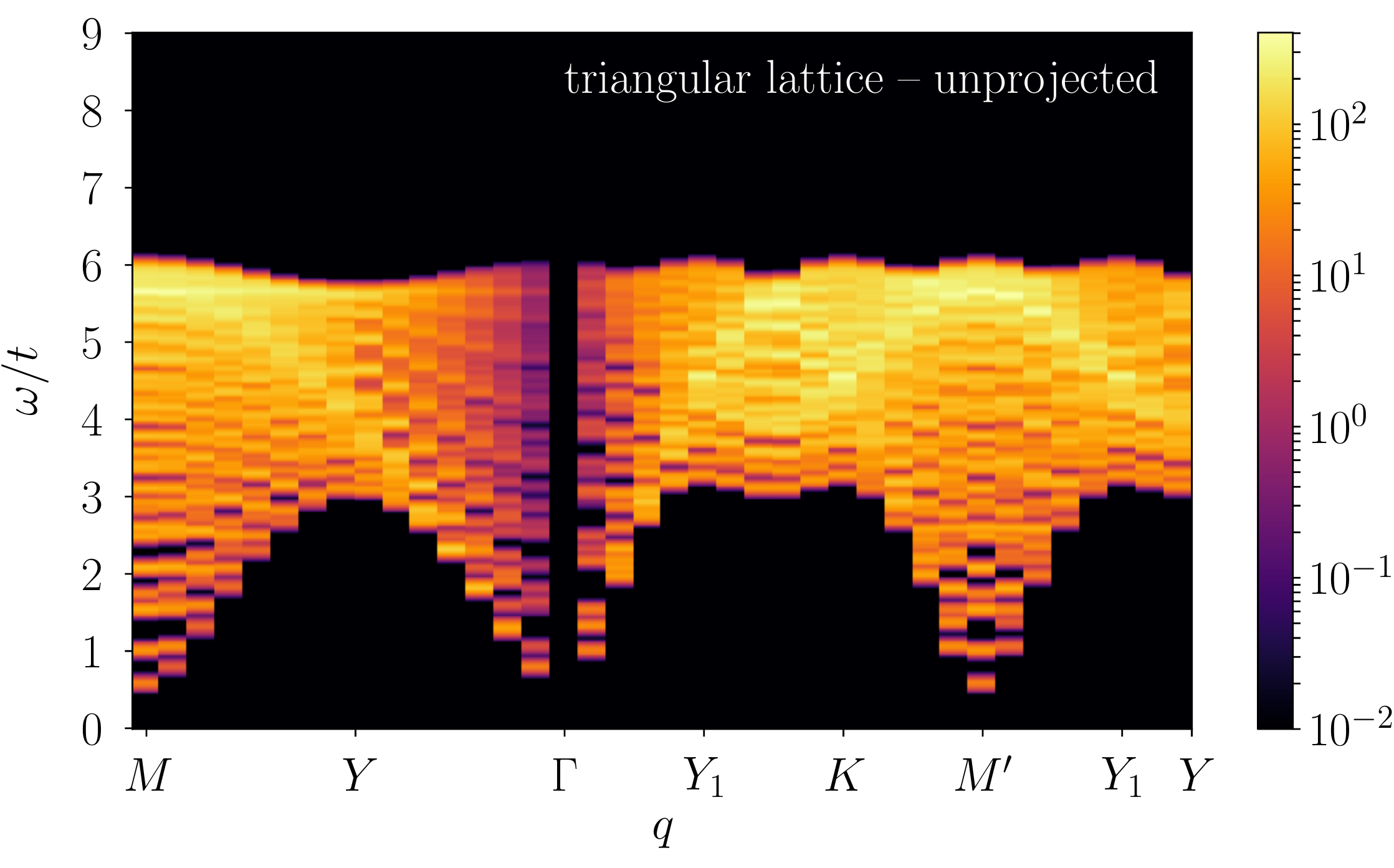}
\caption{\label{fig:125}
The dynamical structure factor for the $J_1-J_2$ Heisenberg model on the $30 \times 30$ cluster with $J_2/J_1=0.125$. The variational results 
(left panel) are compared to the ones obtained from the unprojected Abrikosov fermion Hamiltonian $\mathcal{H}_0$ of Eq.~(\ref{eq:auxham}) with 
$t=1$ and $h=0$ (right panel). The path along the Brillouin zone is shown in Fig.~\ref{fig:latt}. We applied a Gaussian broadening of $\sigma=0.02J_1$ 
to the variational results. Notice that, for the unprojected data, the energy scale is given by the hopping amplitude $t$ of the unprojected 
Hamiltonian~(\ref{eq:auxham}), instead of $J_1$. In addition, the broadening has been rescaled in order to account for the larger bandwidth of the 
spectrum.}
\end{figure*}

\begin{figure*}
\includegraphics[width=\columnwidth]{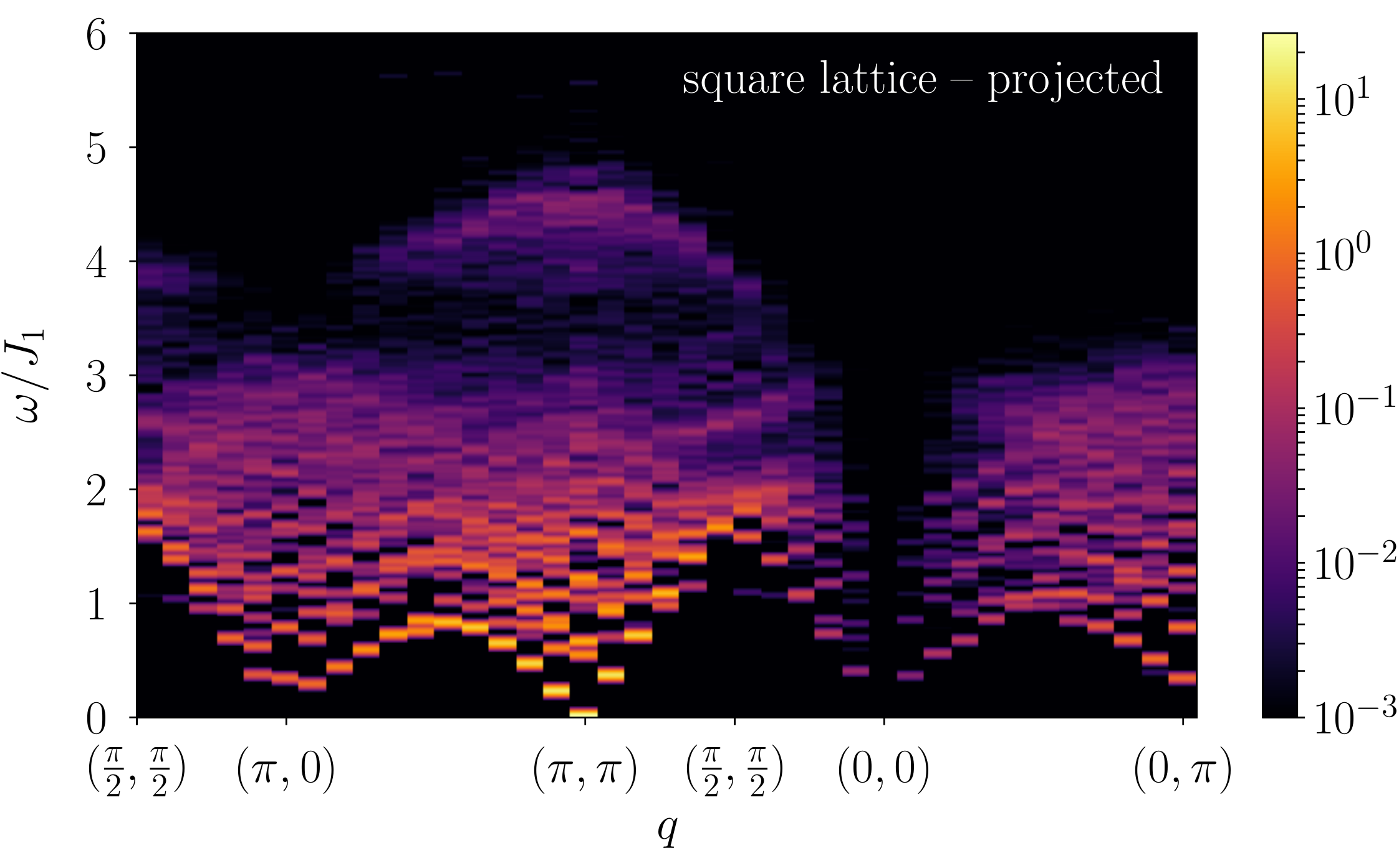}\hfill
\includegraphics[width=\columnwidth]{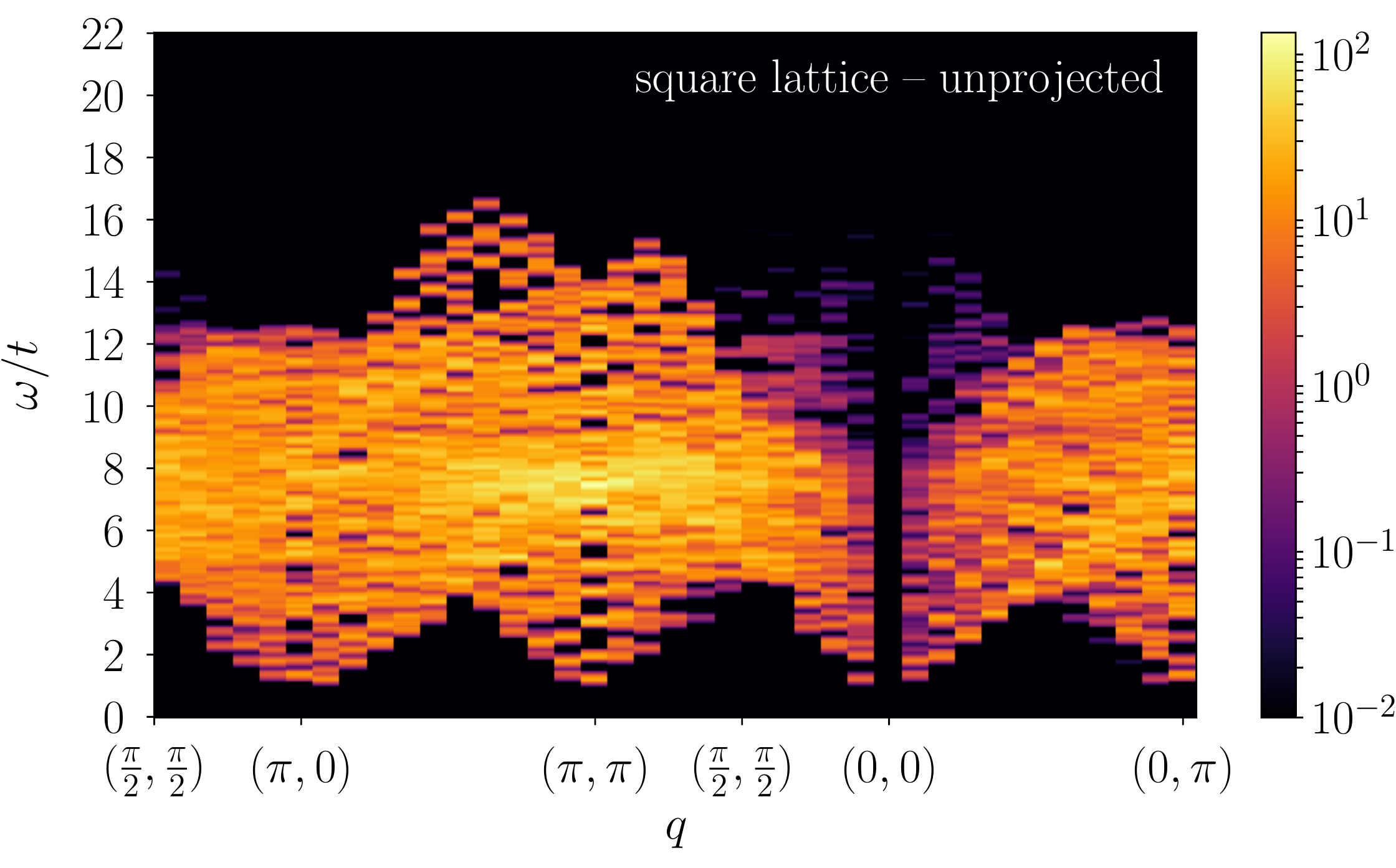}
\caption{\label{fig:square}
The dynamical structure factor for the $J_1-J_2$ Heisenberg model on the square lattice ($22 \times 22$) with $J_2/J_1=0.55$. The 
variational results (left panel) are compared to the ones obtained from the unprojected Abrikosov fermion Hamiltonian $\mathcal{H}_0$ 
(right panel), which contains a flux-phase hopping (of strength $t$) and a $d_{xy}$ pairing (see Ref.~\cite{ferrari2018b} for details). 
We applied a Gaussian broadening of $\sigma=0.02J_1$ to the variational results. Notice that, for the unprojected data, the energy scale 
is given by the hopping amplitude $t$ of the unprojected Hamiltonian of Ref.~\cite{ferrari2018b}, instead of $J_1$. In addition, the broadening 
has been rescaled in order to account for the larger bandwidth of the spectrum.}
\end{figure*}


A much more realistic spectrum is obtained when considering a finite fermion hopping $t$ (with the $\pi$-flux pattern shown in 
Fig.~\ref{fig:latt}), as well as the optimized value of the fictitious magnetic field $h$ (and the Jastrow factor). The results for
the $30 \times 30$ lattice are reported in Fig.~\ref{fig:AF120}. In this case, there are several excitations with a finite weight for
each momentum, thus reproducing the existence of a broad continuum, which extends up to relatively large energies. We would like to
mention that, with respect to the square lattice~\cite{ferrari2018b,dallapiazza2015,yu2018}, here many more excitations for each momentum
possess a visible spectral weight. Within this calculation, we identify the lowest-energy excitation $E_{0}^{q}$ as the magnon peak. This
assumption is corroborated by the results shown in Fig.~\ref{fig:dispersions}, where the variational energies $E_{0}^{q}$ closely follow 
the magnon branch obtained by series expansions. Instead, identifying the lowest-energy peak as the bottom of the continuum is not very 
plausible, since a much broader signal should be present in this case. In this regard, the basis set that is used here for the excited 
states is made of particle-hole spinon excitations on top of the ground state of the auxiliary Hamiltonian of Eq.~(\ref{eq:auxham}), before 
Gutzwiller projection. For this reason, we argue that, in general, our approach is particularly suited to capture (i) two-spinon excitations 
or (ii) bound states of spinons, e.g., magnons. Multi-magnon excitations are expected to show up with a reduced intensity. In order to discuss 
the issue of magnon decay, we apply a kinematic argument (as done both in the linear spin-wave approach~\cite{chernyshev2006,chernyshev2009} 
and within DMRG~\cite{verresen2018}) and we consider all the possible two-magnons decays, which fulfill the conservation of momenta, i.e., 
$E_{q}= \min_k \{ E_{0}^{q-k}+E_{0}^{k}\}$. For this purpose, we computed the spectrum $E_{0}^{k}$ for all the $k$-vectors in the Brillouin 
zone on the $30 \times 30$ lattice. The outcome is that the bottom of the two-magnon continuum, defined by the kinematic analysis, lies above 
the magnon branch. These results clearly indicate an avoided decay in a large part of the Brilloiun zone, as suggested by DMRG calculations,
which considered certain (high-energy) parts of the magnon dispersion~\cite{verresen2018}. Still, we cannot exclude the existence of small 
regions where the magnon decay may persist, especially close to the gapless points. In this respect, within the linear spin-wave approach, 
the different velocities of the excitation spectrum at $\Gamma$ and $K$ immediately lead to an unstable magnon branch close to the $\Gamma$ 
point~\cite{chernyshev2006,chernyshev2009}. Should this aspect be a genuine feature of the model, the magnon would be unstable in a small 
part around the center of the Brillouin zone. Unfortunately, given the finiteness of the cluster used in our numerical calculations, we cannot 
reliably estimate the slope of the magnon spectrum at $\Gamma$ and $K$ and, therefore, make definitive statements for this issue.

Here, we would like to notice the strong renormalization of the magnon branch with respect to spin-wave calculations, see 
Fig.~\ref{fig:dispersions}. Most importantly, we emphasize that, within this most accurate calculation, the magnon branch shows a roton-like 
minimum not only at $M$, but also at $Y_1$, i.e., the midpoint of the edge of the magnetic Brillouin zone (see also Fig.~\ref{fig:dispersion}), 
as already detected by neutron scattering measurements in Ba$_3$CoSb$_2$O$_9$~\cite{ito2017}. This feature was not captured by the previous
series expansion calculations~\cite{zheng2006} but, instead, has been observed also by recent DMRG calculations on an infinitely long cylinder 
(with a small circumference $L=6$)~\cite{verresen2018} and has been interpreted as the hallmark for the absence of magnon decay. In order to 
make a closer comparison with DMRG data, we perform the variational calculations on a long cylinder ($84 \times 6$) along the same path in the 
Brillouin zone as the one that has been considered in Ref.~\cite{verresen2018}. The results are shown in Fig.~\ref{fig:cylin}. Here, the large 
number of lattice points along the cylinder allows us to have a detailed resolution of the magnon branch, which closely follows the one obtained 
by DMRG. In particular, we can estimate the bottom of the continuum by evaluating $E_{q}=\min \{ E_{0}^{q-K}+E_{0}^{K}, E_{0}^{q+K}+E_{0}^{-K} \}$, 
where we consider the possible decays involving a magnon at $K$ and $-K$. In doing so, we find that the lowest-energy excitation $E_{0}^{q}$ is 
always below $E_{q}$, indicating that well defined branch exists and magnon decay is avoided. We finally remark that a roton minimum is detected 
along the same path as the one studied by Verresen and collaborators~\cite{verresen2018}, strongly suggesting that this is a genuine feature of 
the Heisenberg model.

\subsection{The $J_1-J_2$ model}
 
We now move to the case where also a next-nearest-neighbor coupling $J_2$ is present. Within our variational approach, a gapless
spin-liquid phase is stabilized for $0.08 \lesssim J_2/J_1 \lesssim 0.16$; here, the fictitious magnetic field vanishes in the
thermodynamic limit and the best wave function only contains fermionic hopping (with $\pi$-flux threading half of the triangular
plaquettes)~\cite{iqbal2016}. On a finite size, a small value of $h$ can be stabilized, as well as a tiny Jastrow pseudopotential.
Still, we verified that these ingredients do not cause sensible differences in the dynamical structure factor. In Fig.~\ref{fig:007},
we show the results for the $30 \times 30$ cluster and for two values of $J_2/J_1$, which are very close to the transition point, one 
still inside the magnetic phase ($J_2/J_1=0.07$) the other one in the spin-liquid region ($J_2/J_1=0.09$). By approaching the quantum 
phase transition, the major modification of the spectrum comes from the softening of the magnon excitation at the $M$ points. This 
feature closely resembles the case of the frustrated $J_1-J_2$ model on the square lattice, previously studied with the same numerical 
technique~\cite{ferrari2018b}, where a softening is clearly detected for ${\bf q}=(\pi,0)$ [and $(0,\pi)$]. In this latter case, this 
fact has been connected to the progressive deconfinement of spinons that have gapless (Dirac) points at ${\bf q}=(\pm \pi/2,\pm \pi/2)$. 
We would like to mention that the possibility to have (gapped) almost-deconfined spinon in the unfrustrated Heisenberg model has been 
suggested by a recent quantum Monte Carlo calculation~\cite{shao2017}; moreover, clear signatures for deconfined spinons at the transition 
between an antiferromagnetically ordered phase and a valence-bond crystal have been reported in the so-called $J-Q$ model~\cite{ma2018}. 
On the triangular lattice, the softening of the spectrum at the $M$ points is a direct consequence of the Dirac points at 
${\bf q}=(0,\pm \pi/\sqrt{3})$ in the spinon band structure. Therefore, we expect both $M$ and $K$ points to be gapless at the transition 
(as well as $Y_1$, which can be obtained by combining $M$ and $K$ vectors). Indeed, this is necessary for a continuous phase transition, 
as the one that appears in the $J_1-J_2$ Heisenberg model, according to ground-state calculations~\cite{iqbal2016}.

In Fig.~\ref{fig:125}, we report the dynamical structure factor for $J_2/J_1=0.125$. The 
spin-liquid state is characterized by a broad continuum that extends up to relatively large energies. In particular, around the $M$ 
points, the magnon roton-like minima of the ordered phase fractionalize into an incoherent set of excitations at low energies. This 
feature is compatible with the existence of Dirac points in the unprojected spectrum of the auxiliary Hamiltonian $\mathcal{H}_0$, 
see Fig.~\ref{fig:125}. By contrast, a strong signal in the lowest-energy part of the spectrum is detected around the $K$ points, where 
the unprojected spinon spectrum is instead gapped. In this respect, the Gutzwiller projection is fundamental to include interaction
among spinons in a non-perturbative way and give a drastic modification of the low-energy features. This is a distinctive aspect of 
the triangular lattice, since, on the square lattice, all the low-energy (gapless) points observed in presence of the Gutzwiller 
projector [i.e. ${\bf q}=(0,0)$, $(\pi,\pi)$, $(\pi,0)$ and~$(0,\pi)$] already exist in the non-interacting picture~\cite{hu2013}, 
see Fig.~\ref{fig:square}. We would like to emphasize that, in contrast to the magnetically ordered phase, where no visible spectral 
weight is present right above the magnon branch (see Fig.~\ref{fig:AF120}), in the spin-liquid phase the continuum is not separated 
from the lowest-energy excitation. This outcome corroborates the fact of having deconfined spinons in the magnetically disordered phase. 
The intense signal at $K$ points immediately implies strong (but short-range) antiferromagnetic correlations in the variational wave 
function, which are absent in the unprojected $\pi$-flux state (by contrast, on the square lattice, the $\pi$-flux state has already 
significant antiferromagnetic correlations built in it).

The presence of low-energy spectral weight at the corners of the Brillouin zone could be ascribed to the existence of critical monopole 
excitations, as suggested by the analysis of Ref.~\cite{song2018}. In fact, the Gutzwiller projector, which imposes single occupacy
on each lattice site, introduces temporal fluctuations of the gauge fields that are completely frozen within the non-interacting
picture (i.e., within the unprojected wave function). Even though we cannot exclude a more conventional picture where a bound state of 
spinons is responsible for the intense signal around $K$, it is plausible that this feature originates from the existence of gauge
fields, which emerge in the field-theoretical description of spin liquids~\cite{savary2016}. While gauge fields are known to predominantly
contribute to spectral functions of specific Kitaev spin liquids with $\mathcal{Z}_2$ magnetic fluxes~\cite{knolle2014}, our calculations 
suggest that monopole excitations may give some relevant signature in the spin-liquid phase of the $J_1-J_2$ Heisenberg model on the 
triangular lattice. Remarkably, on the $30 \times 30$ cluster, the lowest-energy excitation at $K$ is slightly higher inside the 
spin-liquid phase (i.e., for $J_2/J_1=0.125$) than close to the critical point (i.e., for $J_2/J_1 \approx 0.08$), see Figs.~\ref{fig:007}
and~\ref{fig:125}. This fact may suggest the possibility that this kind of excitation may be slightly gapped in the spin-liquid region, 
while being gapless at the critical point. We finally highlight the existence of an unexpected high-energy dispersing mode, which bends 
from the $\Gamma$ point down into the continuum, being seemingly connected to the low-energy excitation at $K$. A comparison with other 
numerical techniques will be needed to clarify whether this feature is a genuine aspect of the model or an artifact of the present 
variational approach.

\section{Conclusions}

In this work, we performed variational Monte Carlo calculations to estimate the dynamical structure factor of the $J_1-J_2$
Heisenberg model on the triangular lattice. The results for $J_2=0$ are consistent with the existence of a well-defined magnon branch
in the whole Brillouin zone, in agreement with recent DMRG calculations~\cite{verresen2018}. This outcome contrasts the 
semiclassical predictions~\cite{chernyshev2006,chernyshev2009}, which suggested the presence of magnon decay in a large portion 
of the Brillouin zone. When a finite $J_2$ super-exchange is included and the spin-liquid phase is approached, a clear softening of
the spectrum is detected around the $M$ points, in close similarity to what happens on the square lattice~\cite{ferrari2018b}. 
Remarkably, the low-energy physics of the spin liquid phase cannot be fully described by the unprojected spinon picture, since, 
besides gapless excitations at $M$ and $M^\prime$, there are anomalously low-energy states appearing around the $K$ points.  
Our numerical calculations provide an indisputable evidence of the fact that the non-interacting (i.e., unprojected) spinon spectrum
is not sufficient to fully explain the low-energy spectrum detected by the dynamical structure factor. In light of the recent 
field-theoretical analysis~\cite{song2018}, the natural interpretation of the spectral features around the corners of the Brillouin
zone comes from the existence of low-energy monopole excitations. This outcome is particularly important, since it would give a 
direct signature of the fact that these theoretical approaches correctly capture the nature of the spin-liquid phase. 
We hope that the present results will motivate future investigations in this direction.

\acknowledgements
We are particularly indebted to T. Li, for pointing out interesting aspects of the problem, and A. Chernyshev, for highlighting some 
aspects of his results. We also acknowledge C. Batista, Y.-C. He, A. Parola, F. Pollmann, R. Verresen, and A. Vishwanath for useful 
discussions.

\bibliographystyle{apsrev4-1}

\end{document}